\documentclass[10pt]{iopart}
\pdfoutput=1
\usepackage[english]{babel}   
\usepackage{graphicx}
\usepackage{braket}
\usepackage{url}
\usepackage{nicefrac}
\usepackage[utf8]{inputenc}

\usepackage{siunitx}
\DeclareSIUnit\erec{\text{\ensuremath {\text{E}_{\textup {rec}}}}}
\DeclareSIUnit\deg{\degree}
\DeclareSIUnit\nph{\text{n}_\text{ph}}
\DeclareSIUnit\cts{\text{Cts}}

\def \t{\text}

\begin{document}

\title[]{Continuous feedback on a quantum gas coupled to an optical cavity}
\author{Katrin Kroeger$^1$, Nishant Dogra$^{1,2}$, Rodrigo Rosa-Medina$^1$, Marcin Paluch$^1$, Francesco Ferri$^1$, Tobias Donner$^{1,*}$, Tilman Esslinger$^1$}
\address{$^1$ Institute for Quantum Electronics, ETH Zurich, 8093 Zurich, Switzerland \\
$^2$ present address: Cavendish Laboratory, University of Cambridge, J. J. Thomson Avenue, Cambridge CB3 0HE, United Kingdom}
\ead{donner@phys.ethz.ch}

\begin{abstract}
We present an active feedback scheme acting continuously on the state of a quantum gas dispersively coupled to a high-finesse optical cavity. The quantum gas is subject to a transverse pump laser field inducing a self-organization phase transition, where the gas acquires a density modulation and photons are scattered into the resonator. Photons leaking from the cavity allow for a real-time and non-destructive readout of the system. We stabilize the mean intra-cavity photon number through a micro-processor controlled feedback architecture acting on the intensity of the transverse pump field. The feedback scheme can keep the mean intra-cavity photon number $n_\t{ph}$ constant, in a range between $n_\t{ph}=\num{0.17\pm 0.04}$ and $n_\t{ph}=\num{27.6\pm 0.5}$, and for up to \SI{4}{\second}. Thus we can engage the stabilization in a regime where the system is very close to criticality as well as deep in the self-organized phase. The presented scheme allows us to approach the self-organization phase transition in a highly controlled manner and is a first step on the path towards the realization of many-body phases driven by tailored feedback mechanisms.
\end{abstract}

\maketitle

\section{Feedback on quantum gases}
Ultracold atomic quantum gases are a well-suited platform to study transitions and crossovers between different phases of matter. Prominent examples are the phase transition from a thermal gas to a Bose-Einstein condensate \cite{anderson_observation_1995,davis_bose-einstein_1995} and between a superfluid and a Mott insulator \cite{greiner_quantum_2002}, or the crossover between a Bose-Einstein condensate of molecules and a Bardeen-Cooper-Schrieffer superfluid of loosely bound pairs in quantum degenerate Fermi gases \cite{bourdel_experimental_2004,bartenstein_crossover_2004,greiner_probing_2005}. Another well studied example is the transition to a superradiant or self-organized phase in the driven-dissipative Dicke model \cite{baumann_dicke_2010}. 

A characteristic of cold atom experiments is that most probing techniques are inherently destructive \cite{Ketterle_making_1999}. This requires experiments to be repeated many times with the same initial conditions, such that statistically significant findings can be derived. The successive preparation of the system with identical parameters proves challenging and therefore often requires postselection or additional procedures during state preparation. In a recent experiment, non-destructive Faraday imaging was employed to measure the atom number and adjust the subsequent optical evaporation to prepare a number stabilized ultracold atomic cloud \cite{gajdacz_preparation_2016}. In other experiments, the transmission spectrum of a cavity was monitored to trigger the start of the experiment once a set atom number, detected via the dispersive shift of the cavity resonance, was reached \cite{haas_entangled_2014,zhiqiang_nonequilibrium_2017}.

Coupling an atomic gas to a cavity comes with a non-destructive measurement channel. The continuous photon leakage through the cavity mirrors can be used to monitor the evolution of the system in real time. Starting point for our feedback scheme is an experimental setup in which we prepare a degenerate Bose gas inside a high-finesse optical cavity. By tuning the strength of an external drive field, the combined atom-cavity system is undergoing a self-organization phase transition which can be mapped onto the Dicke model \cite{baumann_dicke_2010}. The intra-cavity light field is related to the order parameter of this phase transition, providing in principle real-time access to critical properties such as the fluctuations of the order parameter. However, since the system close to the critical point is highly susceptible to minute drifts of any parameter, such a measurement is challenging and requires many experimental runs and a sophisticated data analysis \cite{brennecke_real-time_2013,landig_measuring_2015}. Here, we stabilize the state of the system via detecting the photons leaking through the cavity mirrors and applying an according feedback signal to the control parameter. This way, the lifetime of the self-organized phase close to the phase transition can be dramatically extended. 

Going a step further, feedback could be used to control phase transitions and engineer novel non-equilibrium phases in driven many-body systems. The combination of feedback and weak measurements promises to realize new feedback-induced phase transitions and control of their critical exponent \cite{ivanov_feedback-induced_2019}. The simulations of interesting many-body problems like spin-bath models, Ising type interactions, the Lipkin-Meshkov-Glick model and Floquet time crystals through specifically engineered feedback schemes have also been proposed \cite{ivanov_feedback-induced_2019}. Applying time-delayed feedback to the driven-dissipative Dicke model opens the prospect to study non-equilibrium dynamics with fixed points and limit cycles in the superradiant regime \cite{kopylov_time-delayed_2015}. 

In this work, we report on the basic building block to implement such feedback schemes and present a feedback architecture designed to stabilize the mean intra-cavity photon number. 
We first review the self-organization phase transition in our system. Following, we present the technical implementation of our feedback scheme. We then compare experimental realizations with and without the feedback and demonstrate stabilization of the mean intra-cavity photon number in a wide range.

\begin{figure}[b]
	\centering
	\includegraphics[width=0.6\textwidth]{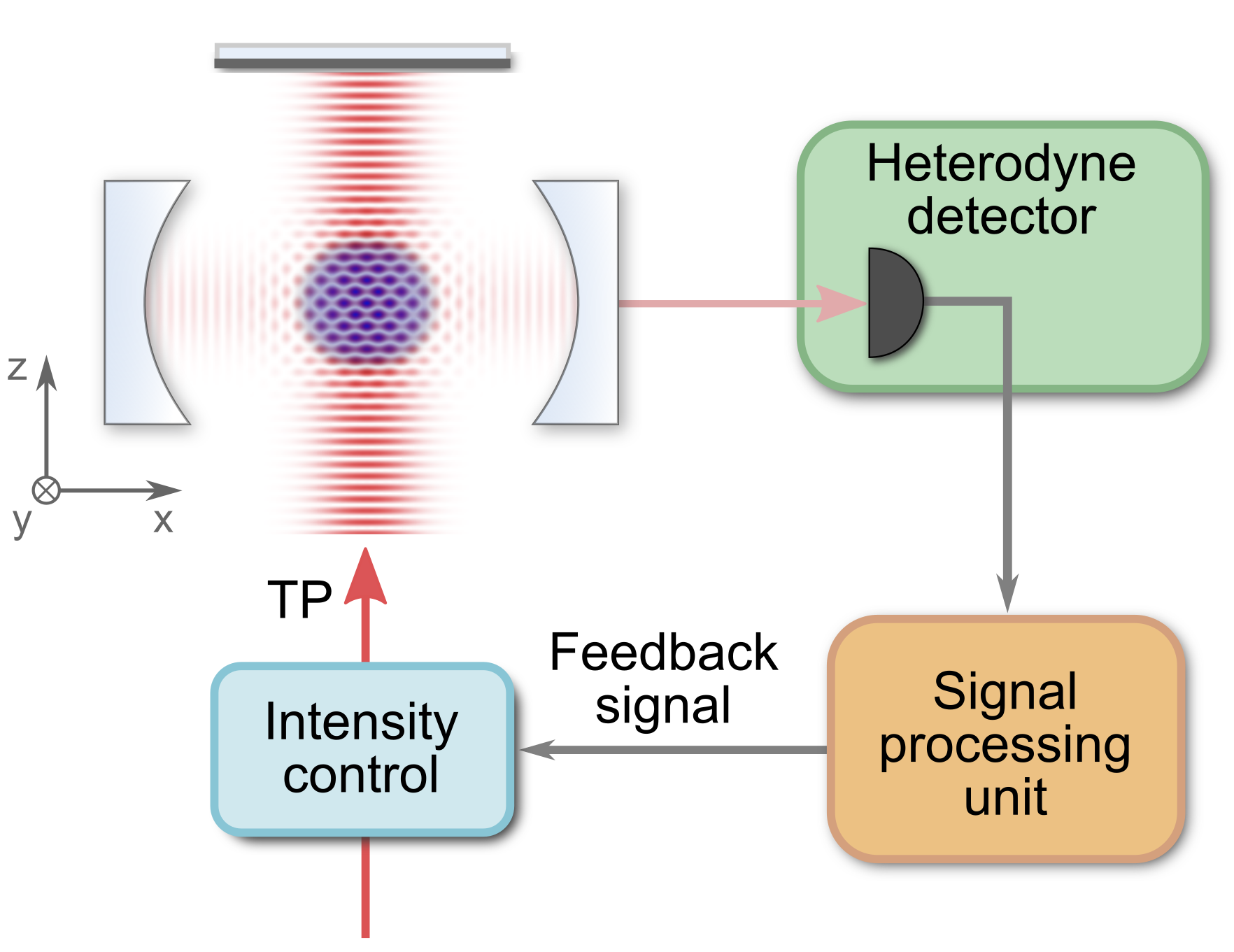}
	\caption{Sketch of the experimental setup. An atomic gas (blue) inside an optical cavity is illuminated with a retroreflected standing wave transverse laser pump (TP). The system undergoes a self-organization phase transition for a sufficiently high intensity of the TP. In the self-organized phase, the atomic density is modulated and an intra-cavity light field is built up. The intra-cavity light field leaks through the cavity mirrors and is recorded by a heterodyne detector (green box). The signal processing unit calculates the present mean intra-cavity photon number and the feedback signal required to stabilize it to a set value (orange box). The feedback signal is sent to the intensity control of the TP (blue box). When the feedback is not enabled, an external control signal is sent to the intensity control of the TP.}
	\label{fig:fig1}
\end{figure}

\section{Self-organization phase transition of an atomic gas in a cavity}
We prepare a $^{87}$Rb Bose-Einstein condensate (BEC) inside a high-finesse optical cavity and illuminate it with a retroreflected standing wave transverse laser pump (TP). The frequency of the TP is red detuned from atomic resonance. Both the exact detuning between the TP frequency and the cavity resonance frequency as well as the intensity of the TP are adjustable. The intensity of the TP controls the depth of the generated standing wave lattice. The setup is sketched in figure \ref{fig:fig1}.

At a critical lattice depth of the transverse pump the system undergoes a phase transition to a self-organized state, which can be mapped to the Dicke model \cite{baumann_dicke_2010}. A density modulation of the atomic cloud in the form of a chequerboard pattern emerges in the self-organized phase. Photons from the TP are coherently scattered off this density modulation into the cavity mode via Bragg scattering. In this way, a coherent intra-cavity light field is built up. The finite critical pump lattice depth required for this phase transition results from the competition between kinetic and potential energy. The kinetic energy cost increases as the density modulation of the gas becomes more pronounced. At the same time, the potential energy is lowered due to the light shift experienced by the atoms in the interference lattice between TP and intra-cavity light field. 

The intra-cavity light field continuously leaks through the cavity mirrors due to the finite cavity decay rate. We detect the photons leaking out of the cavity with a heterodyne detector, which enables the frequency-resolved reconstruction of both electric field amplitude and phase within its detection bandwidth.

We can infer the phase transition from the normal to the self-organized state from two signatures. The first signature is observed in absorption images of the atoms, which are taken after suddenly switching off all light fields and subsequent ballistic expansion of the atomic cloud. For a sufficiently long expansion time, the atomic image reveals the momentum distribution of the atomic cloud. The occurence of momentum peaks corresponding to a chequerboard modulation of the atomic density signals that the atoms were self-organized. This measurement is destructive. The second signature is the onset of a coherent intra-cavity light field, which is continuously recorded via the heterodyne detector. 

When the TP lattice depth is ramped up to a constant value above the critical lattice depth, the mean intra-cavity photon number decreases over time. This decline is caused by the presence of the light fields which leads to heating of the atomic cloud through incoherent scattering and subsequent atom loss. A decrease in atom number results in a lower collective atom-field coupling strength and leads therefore to less intra-cavity photons for a constant TP lattice depth \cite{baumann_dicke_2010, SI}. This effect is especially pronounced for high mean intra-cavity photon numbers, where the system is deep in the self-organized phase. Furthermore, when the system just crossed the phase transition point, the atomic modulation is weak and the mean intra-cavity photon number low. As the system behaves non-linearly close to criticality \cite{dimer_proposed_2007}, it is highly susceptible to any drifts. This makes it very challenging to study the system at low intra-cavity photon numbers in the vicinity of the self-organization phase transition.

To overcome these challenges, we designed a microcontroller based feedback architecture which stabilizes the mean intra-cavity photon number. The idea is schematically shown in figure \ref{fig:fig1}. We make use of the non-destructive and continuous readout of the system's state provided by the cavity. The signal processing unit calculates the current mean photon number from the signal recorded by the heterodyne detector. It determines the deviation from the desired mean photon number and calculates the feedback signal required to stabilize the system. The feedback signal is sent to the intensity control of the TP laser field.

\section{Heterodyne detection and feedback circuit}
\begin{figure}[t!]
	\centering
	\includegraphics[width=0.8\textwidth]{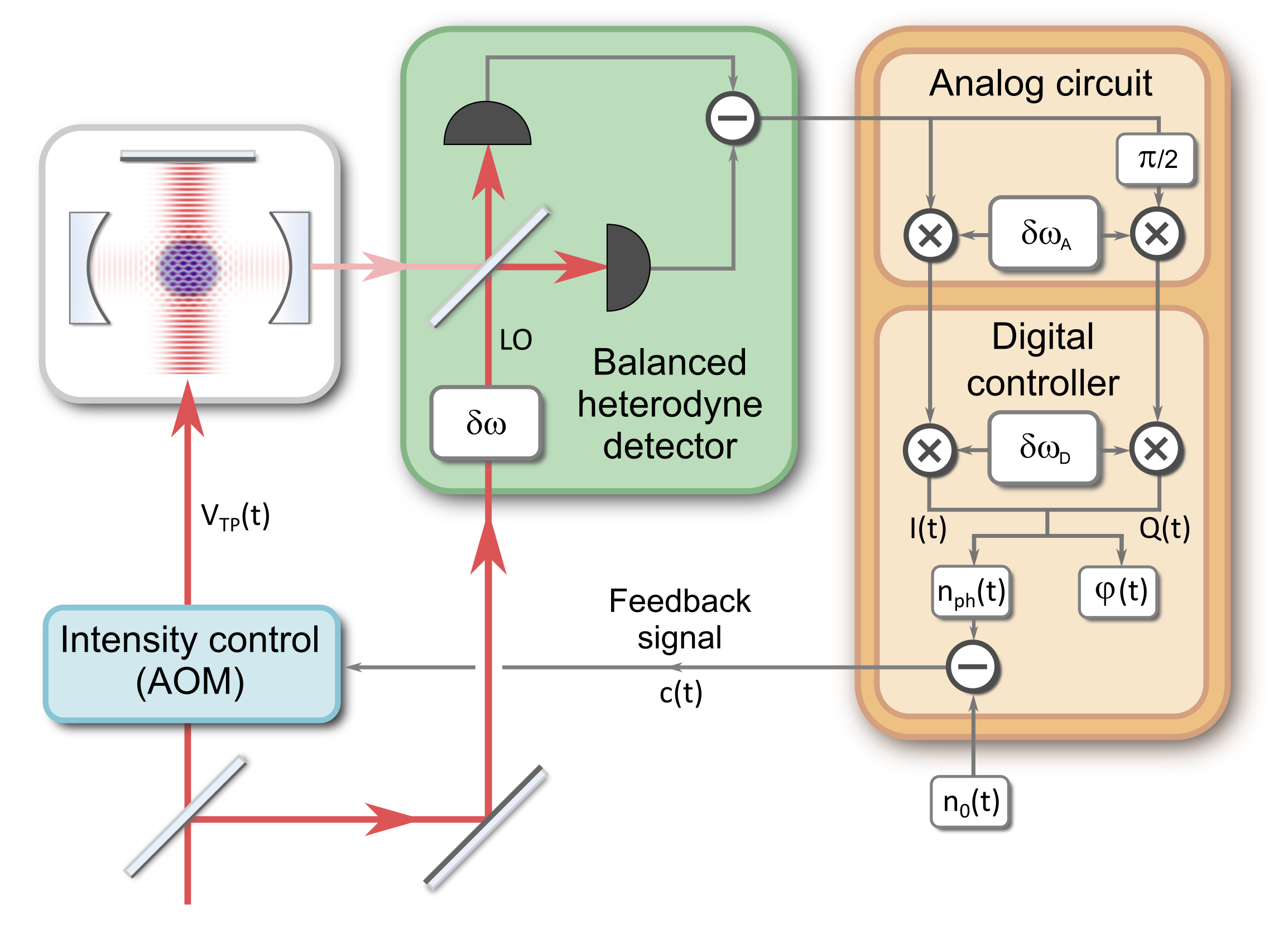}
	\caption{Implementation of the feedback scheme. The intra-cavity light field leaking through the mirror is interfered with a local oscillator and detected by a balanced heterodyne detector (green box). The quadratures are first down-mixed in an analog circuit and then further processed in a digital controller to obtain the time-dependent quadratures $I(t)$, $Q(t)$ and the mean intra-cavity photon number $n_\t{ph}(t)$ of the intra-cavity light field. The phase $\varphi(t)$ of the intra-cavity light field can also be accessed. The determined photon number is compared with the set value $n_{0}(t)$ of the mean intra-cavity photon number (orange box). The calculated feedback signal $c(t)$ is sent to the intensity regulation of an acousto-optic modulator (AOM) in the TP path, changing the value of the pump lattice depth $V_\t{TP}(t)$ (blue box). Frequencies are adjustable and set to $\delta \omega =-2\pi\cdot\SI{59.55}{\mega\hertz}$, $\delta \omega_A =2\pi \cdot \SI{59.503}{\mega\hertz}$ and $\delta \omega_D=2\pi\cdot\SI{47}{\kilo\hertz}$ for the measurements presented.}
	\label{fig:fig2}
\end{figure}

The intra-cavity light field is detected in a balanced heterodyne measurement \cite{baumann_exploring_2011}. Figure \ref{fig:fig2} depicts the details of the optical heterodyne measurement, the signal processing stages and the feedback circuit.
The frequency of the intra-cavity light field in the self-organized phase equals the frequency of the TP because photons from the TP are scattered off a static atomic Bragg grating into the cavity mode. This Bragg grating is created in a self-consistent way as a result of the chequerboard potential provided by the interfering light fields.
The light field leaking through one of the cavity mirrors is interfered with a local oscillator laser beam (LO) on a beam splitter and sent to two balanced photodiodes. The frequency of the LO is offset by $\delta \omega =- 2\pi\cdot\SI{59.55}{\mega\hertz}$ from the TP frequency. The difference signal of the two photodiodes is split, partly phase-shifted and subsequently down-mixed in an analog mixer with a frequency of $\delta \omega_A =2\pi\cdot\SI{59.503}{\mega\hertz}$. These two quadratures are further digitally down-mixed with a variable frequency $\delta \omega_D$ in the digital controller. The resulting quadratures are referred to as $I(t)$ and $Q(t)$.
We use $\delta \omega_D =2\pi\cdot\SI{47}{\kilo\hertz}$ to obtain information about the light field stemming from the self-organization process. We calculate the time-dependent mean intra-cavity photon number $n_\t{ph}(t)$ of the intra-cavity light field from the quadratures $I(t)$ and $Q(t)$. The error signal for the feedback algorithm is calculated as the difference between the set mean photon number $n_{0}(t)$ and the measured mean photon number $n_\t{ph}(t)$:
\begin{eqnarray}
e(t)=n_0(t)-n_\t{ph}(t).
\end{eqnarray}
The feedback signal $c(t)$ is determined through the formula
\begin{eqnarray}
c(t)=c(t-\Delta t)+p\cdot e(t),
\end{eqnarray} 
where $\Delta t$ is the temporal distance to the previously sent feedback signal and $p$ is an adjustable gain.

The power of the TP is regulated via the transmission through an acousto-optic modulator controlled by analog electronics. This change in intensity is equivalent to a change in the TP lattice depth $V_\t{TP}(t)$. The setpoint for the TP intensity regulation is either steered externally or dynamically adjusted to the calculated feedback signal $c(t)$. Technical details on the microcontroller and the implemented algorithm can be found in the supplementary material \cite{SI}.

\section{Measurements}
We prepare a nearly pure BEC of $^{87}$Rb with N=\num{63\pm 5e3} atoms in the Zeeman state $\ket{F=1, m_F=-1}$, where $F$ and $m_F$ refer to the total angular momentum manifold and the associated magnetic quantum number. The BEC is confined in a crossed-beam dipole trap centered inside a high-finesse optical cavity. The wavelength of the TP is set to $\lambda_{\t{TP}}=\SI{784.7}{\nano\meter}$. The frequency of the TP is red-detuned by $\Delta_c=-2\pi\cdot\SI{15.011 \pm 0.087}{\mega\hertz}$ from the cavity resonance frequency. This detuning is kept constant in the experiments presented in this work.

\begin{figure}[t!]
	\centering
	\includegraphics[width=0.9\textwidth]{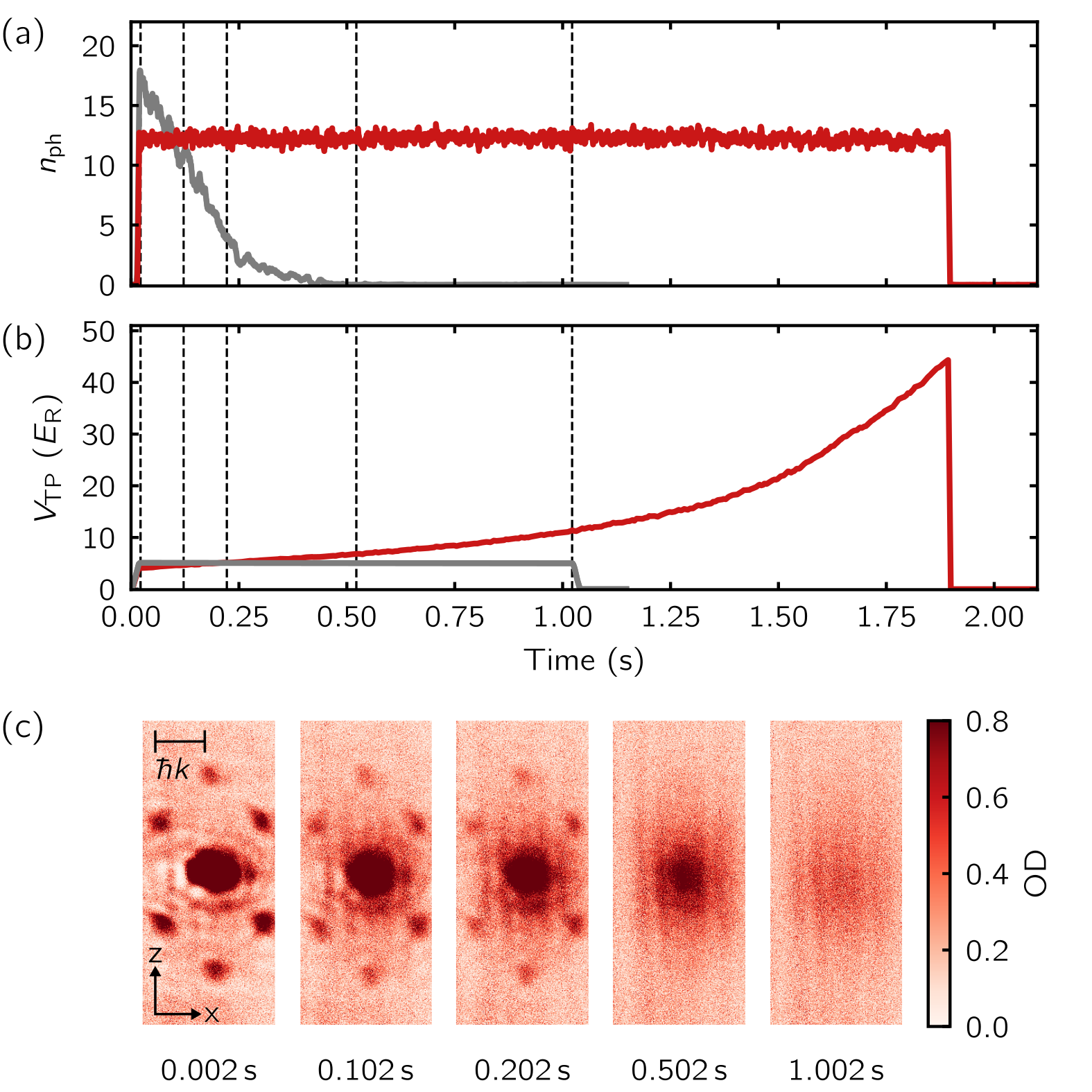}
	\caption{Comparison of self-organization with (red) and without (gray) feedback applied. (a) Mean intra-cavity photon number $n_\t{ph}(t)$ recorded with the heterodyne detector for both scenarios. (b) Lattice depth $V_\t{TP}(t)$ of the transverse pump for the stabilized and non-stabilized case.  Data is displayed using a moving average window of \SI{5}{\milli\second}. (c) Absorption images of the atomic cloud after a sudden switch off of the TP power for different hold times when the system is subject to feedback. The images are saturated to an optical density of OD=0.8 to allow for better visibility of the momentum peaks.  The dashed lines in (a) and (b) indicate the hold times for the images shown in (c).}
	\label{fig:fig3}
\end{figure}

\subsection{Comparison of self-organization with and without feedback}
We compare measurements with and without the active feedback scheme in figure \ref{fig:fig3}. For the measurement without feedback, the ramp of the TP lattice depth follows a predefined protocol: The lattice depth is first increased via an s-shaped ramp \cite{SI} within \SI{20}{\milli\second} to its final value of \SI{5.09\pm 0.06}{\erec}. Here, the recoil energy is defined as $\si{\erec}=\nicefrac{(\hbar k)^2}{2m}$, where $\hbar$ is the reduced Planck's constant, $k=\nicefrac{2\pi}{\lambda_{\t{TP}}}$ is the wavenumber of the TP, and $m$ refers to the atomic mass of $^{87}$Rb. The lattice depth is subsequently held at a constant value for \SI{1}{\second} and finally decreased via an s-shaped ramp within \SI{20}{\milli\second}. The TP lattice depth for this protocol is depicted as the gray curve in figure \ref{fig:fig3}(b). The mean intra-cavity photon number is shown in figure \ref{fig:fig3}(a). It initially reaches $n_\t{ph}=\num{17.76\pm0.08}$ (calculated using a moving average window of \SI{5}{\milli\second}), but decays to half its value within \SI{160}{\milli\second}. There is no signal from self-organization left after about \SI{600}{\milli\second}.

We contrast these results with the measurements with feedback. The TP lattice depth is again increased via an s-shaped ramp within \SI{20}{\milli\second} to \SI{4.45\pm 0.05}{\erec}. The feedback scheme however takes over control as soon as half of the set mean photon number is reached and then steers the ramp of the TP lattice depth. The mean photon number is stabilized to $n_\t{ph}=\num{12.2 \pm 0.3}$ for more than \SI{1.8}{\second}. The sudden decrease in the mean photon number results from switching off the TP power at a maximum lattice depth of \SI{44.5\pm 0.5}{\erec}. This value is the technical upper limit of the TP lattice depth in our feedback scheme due to the dynamic range of the microcontroller. With feedback, the TP lattice depth is increased in a non-linear fashion.

In addition, we show in figure \ref{fig:fig3}(c) absorption images of the atomic cloud for the case with feedback. The images were taken after suddenly switching off the TP after variable hold times. The hold time is defined as the time we let the system evolve after the initial ramp-up stage of \SI{20}{\milli\second}. We obtain the momentum distribution via the absorption images after \SI{8}{\milli\second} of ballistic expansion \cite{Ketterle_making_1999}. 

For short hold times, sharp momentum peaks are visible in the absorption images. We can understand these momentum peaks by invoking a two-mode description of the atomic cloud in momentum space \cite{baumann_dicke_2010}. In this description, the atoms are either occupying the zero momentum mode $\ket{0}=\ket{p_x=0,p_z=0}$ or the excited momentum mode $\ket{k}$ consisting of a symmetric superposition of the four momentum states $\ket{p_x=\pm \hbar k,p_z=\pm \hbar k}$. Here, $p_{x(z)}$ refers to the momentum along the $x$($z$)-direction. The structure of the excited momentum state $\ket{k}$ can be explained by considering the imparted recoil momenta from the two successive scattering events which happen between the atoms and the photons from the TP and cavity light field during self-organization. The momentum peaks at $p_x=0$, $p_z=\pm 2 \hbar k$ stem from the mere presence of the TP lattice. The two-mode description is valid when the atomic gas can be described as a BEC populating mainly the $\ket{0}$ and $\ket{k}$ momentum states. In this regime, the self-organization phase transition can be mapped onto the Dicke model phase transition \cite{baumann_dicke_2010}.

The visibility of the higher momentum peaks decreases for increasing hold times, signaling a loss of coherence. Simultaneously, the thermal fraction increases. This behaviour is due to the TP heating the atomic cloud, leading to the population of higher momentum states not captured by the two-mode description. Eventually we observe self-organization of a density-modulated but completely thermal atomic cloud \cite{black_observation_2003}. The exact theoretical description of the evolution from self-organization in a two-mode system to self-organization in a multi-mode system is the subject of ongoing theoretical efforts \cite{piazza_boseeinstein_2013,kirton_introduction_2019} and beyond the scope of this work.

As described above, the TP lattice depth for the case with feedback is increasing in a non-linear and convex fashion. Qualitatively we comprehend the curve's shape by considering atom loss. The atom loss is compensated with a higher TP lattice depth, which leads to further heating and atom loss, requiring an even higher TP lattice depth. We performed measurements of the atom number for the case with feedback after adiabatically ramping down the TP lattice depth for different hold times up to \SI{1}{\second}, as presented in \cite{SI}. The atom number decay exhibits a linear trend. If we assume such a linear atom number decay during the complete time of the experiment and apply a relation between TP lattice depth and atom number valid in the two-mode description of the atomic cloud, we obtain a similarly convex shaped curve as in figure \ref{fig:fig3}(b).

\begin{figure}[t!]
	\centering
	\includegraphics[width=0.9\textwidth]{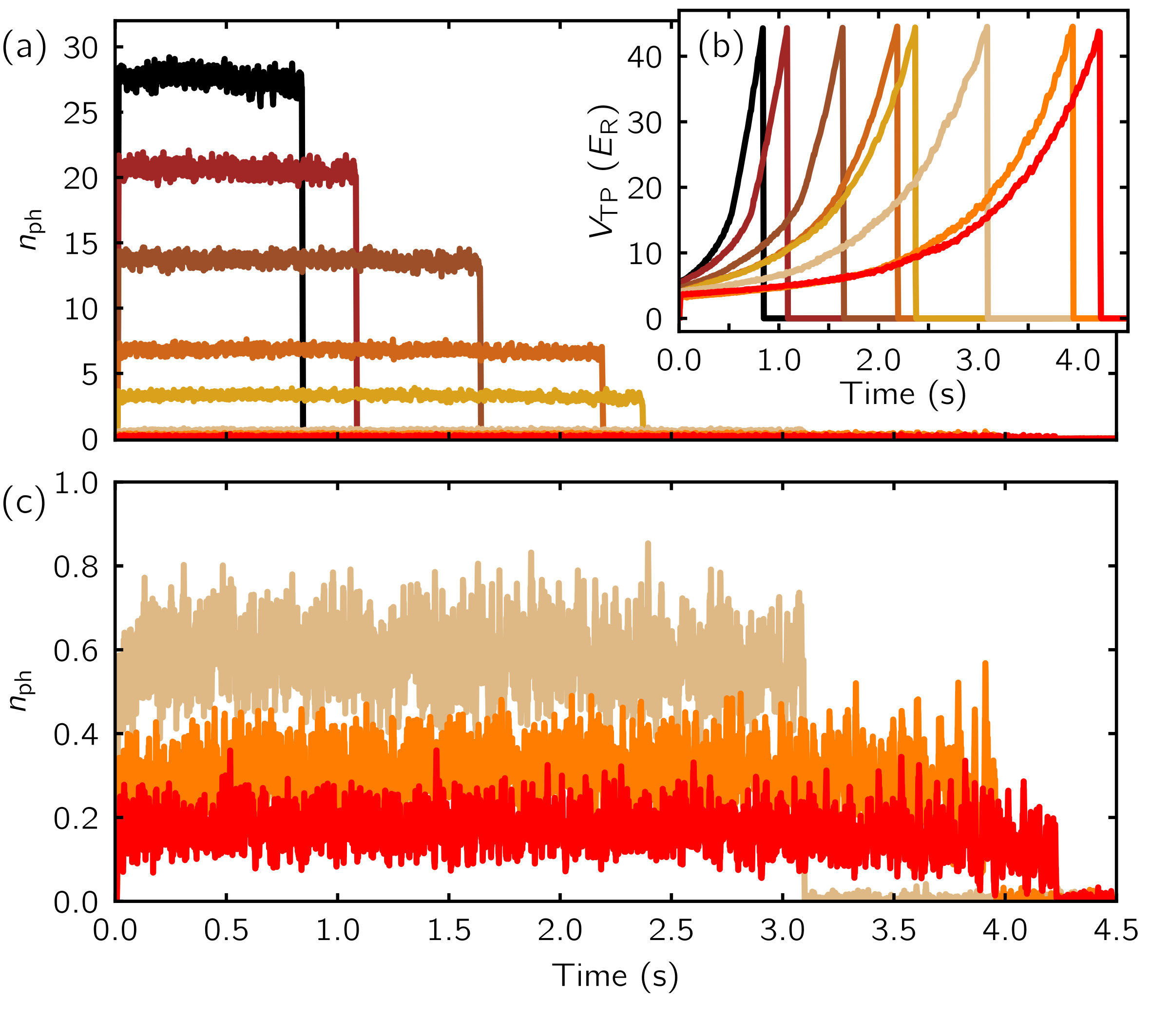}
	\caption{The feedback scheme can stabilize the mean intra-cavity photon number $n_\t{ph}$ in a wide range. This implies that we can stabilize the system both close to the phase transition and deep in the self-organized phase. (a) Constant mean intra-cavity photon number traces of $n_\t{ph}=$ \num{27.6\pm 0.5}, \num{20.6\pm 0.4}, \num{13.7\pm 0.3}, \num{6.8\pm 0.2}, \num{3.28\pm 0.18}, and (zoom in (c)) $n_\t{ph}=$ \num{0.58\pm 0.07}, \num{0.3\pm 0.06}, \num{0.17\pm 0.04}. (b) Corresponding TP lattice depths, as set by the feedback circuit. Data is displayed using a moving average window of \SI{5}{\milli\second}.}
	\label{fig:fig4}
\end{figure}

\subsection{Stabilization on different mean intra-cavity photon numbers}
In an additional set of experiments, we explore the range of mean intra-cavity photon numbers the feedback scheme can stabilize. The data in figure \ref{fig:fig4}(a) and (c) demonstrate that stabilization is possible in a wide range from $n_\t{ph}=\num{0.17\pm 0.04}$ to $n_\t{ph}=\num{27.6\pm 0.5}$. Due to the increased non-linearity close to the phase transition, the stabilization at low photon numbers $n_\t{ph}<1$ requires adjustments: The feedback algorithm takes over control only after the set photon number $n_0$ is reached. In addition, the gain settings both in the feedback software and TP intensity regulation need to be adjusted.

We quantify where the stabilization is engaged relative to the phase transition. For the highest photon number $n_\t{ph}=\num{27.6\pm 0.5}$, the stabilization starts for $V'_{\t{TP}}$ being \SI{50.7\pm 2}{\percent} larger than the critical lattice depth of the self-organization phase transition. Here, $V'_\t{TP}$ is the value of the TP lattice depth for which the desired photon number $n_0$ is reached for the first time. For the lowest achieved photon number $n_\t{ph}=\num{0.17\pm 0.04}$, we start the stabilization at a $V'_\t{TP}$ which is only \SI{1.6\pm1}{\percent} above the critical TP lattice depth, well within the critical regime of this phase transition \cite{brennecke_real-time_2013,landig_measuring_2015}. The respective values for all stabilized $n_\t{ph}$ can be found in \cite{SI}.

\section{Conclusion and outlook}
We developed a microprocessor-based active feedback scheme which stabilizes the number of mean intra-cavity photons during self-organization of a BEC in an optical cavity. The feedback is acting on the intensity of the TP laser beam. The feedback scheme can easily be adjusted to act on other parameters, as for example the frequency detuning between TP and cavity. As the balanced heterodyne detection system records the electric field within the bandwidth of the detection system, the feedback software can be readily modified to calculate other physical quantities, like the phase or information in the spectrum of the light, and control the evolution of these observables. The detection of the cavity field and subsequent regulation can also be extended to the polarization degree of freedom. 

The current capabilities of the feedback scheme enable us to approach the phase transition starting from the self-organized phase in a highly controlled manner. Stabilizing the system close to the phase transition allows to make use of its increased sensitivity with respect to external perturbations for sensing applications \cite{wang_quantum_2014, zanardi_quantum_2008, pezze_adiabatic_2019}. Another interesting prospect is to study fluctuations over long times in the direct vicinity of the phase transition \cite{landig_measuring_2015}.

Instead of stabilizing on a constant mean intra-cavity photon number, the feedback scheme can be modified to modulate the atom-light coupling strength according to more complex schemes. Such control is important to engineer non-equilibrium phases and phase transitions \cite{ivanov_feedback-induced_2019,kopylov_time-delayed_2015, mazzucchi_quantum_2016, grimsmo_rapid_2014}.

\section*{Acknowledgements}
We thank Tobias Delbr{\"u}ck and Chang Gao from the Sensors Group in the Institute of Neuroinformatics, University of Zurich and ETH Zurich, for stimulating discussions and helpful advice. We thank Alexander Frank for electronic support and Fabian Finger for careful reading of the manuscript.
We acknowledge funding from SNF, project numbers 182650 and 175329 (NAQUAS QuantERA), and NCCR QSIT, and funding from EU Horizon2020, ERCadvanced grant TransQ (Project Number 742579).

\newpage
\bibliographystyle{iopart-num}
\bibliography{ms}

\end{document}

\section*{Supplementary Information}

\subsection*{1. Experimental details}
The atoms are held inside a crossed beam dipole trap created by laser fields with wavelength $\lambda_\t{DT}=\SI{852}{\nano\meter}$. The trap frequencies amount to $f_\t{x}=\SI{195\pm 2}{\hertz}$, $f_\t{y}=\SI{40.8\pm 0.3}{\hertz}$ and $f_\t{z}=\SI{119.6\pm0.2}{\hertz}$. \\
The magnetic field points along the negative z-direction as defined in figure~1.\\
The polarization of the TP laser points along the $y$-axis. The lattice depth of the TP is calibrated with Kapitza-Dirac diffraction \cite{gadway_analysis_2009}.\\
The s-shaped ramp used for the TP lattice depth is described by the formula
\begin{align}
V_\t{TP}(t) = V_{\t{TP,start}} + (V_{\t{TP,end}}-V_{\t{TP,start}}) \cdot \left(3\left(\frac{t}{t_0}\right)^2-2\left(\frac{t}{t_0}\right)^3\right),
\end{align}
with the initial (final) lattice depth $V_{\t{TP,start}}$ ($V_{\t{TP,end}}$), the time $t$ and the ramp duration $t_0$.\\
The cavity has a decay rate of $\kappa=2\pi\cdot\SI{1.25}{\mega\hertz}$. It has two birefringent modes, which are tilted by \SI{22}{\deg} with respect to the axes of the $y$-$z$-plane. The detuning $\Delta_c=\omega_p-\omega_{c,y}$ quoted in the text refers to the mainly $y$-polarized mode, with $\omega_c$ being the cavity resonance frequency. The frequency of the mainly y-polarized mode $\omega_\t{c,y}$ is $2\pi\cdot \SI{2.2}{\mega\hertz}$ larger than that of the other cavity mode.

\subsection*{2. Feedback algorithm}
The feedback architecture is implemented on an ARM Cortex-A9 processor of the Xilinx Zynq-7000 System-On-Chip architecture placed on a Zybo Z7-20 board from Digilent. The field programmable gate array was not used for calculations in this project. \\
The voltage from the balanced heterodyne detector is acquired every \SI{3.2}{\micro\second} by an analog-to-digital converter. The mean intra-cavity photon number $n_\t{ph}(t)$ is calculated with an averaging window of $\approx\SI{200}{\micro\second}$ to be less sensitive to noise and fluctuations.\\
The error signal $e(t)$ is the difference between the set mean photon number $n_{0}(t)$ and the measured mean intra-cavity photon number: $e(t)=n_{0}(t)-n_\t{ph}(t)$. The feedback signal $c(t)$ is determined via $c(t)=c(t-\Delta t)+p \cdot e(t)$, where $t$ is the current time, $\Delta t=\SI{6.4}{\micro\second}$ is the time step corresponding to the output sampling rate of the digital-to-analog converter and $p$ is the gain. The overall delay in the feedback signal amounts to $\approx\SI{200}{\micro\second}$ and is dominated by the integration time of the acquired signal.\\

\subsection*{3. Relation between TP lattice depth and atom number}
Within the two-mode description of the BEC, the system can be described by the Hamiltonian of the Dicke model \cite{baumann_dicke_2010}:
\begin{align}
\hat{H}=-\hbar \Delta_c\hat{a}^\dagger\hat{a}+\hbar \omega_0 \hat{J}_z + \frac{\hbar}{\sqrt{N}}\lambda (\hat{a}^\dagger+ \hat{a})\hat{J}_x,
\end{align}
where $\Delta_c$ is the detuning between pump and cavity frequency neglecting the dispersive shift, $\hat{a}$ ($\hat{a}^\dagger$) is the annihilation (creation) operator of the cavity mode in the rotating frame of the transverse pump and $\omega_0(V_\t{TP})$ is the energy difference between the ground and excited atomic momentum state. By considering the zero-momentum state $\ket{0}$ and the excited momentum state $\ket{k}$, as described in the main text, the atomic system is cast into a pseudo spin-1/2 description with corresponding angular momentum operators $\hat{J}_x$ and $\hat{J}_z$. The pseudospin has length $N$, with $N$ being the number of atoms. 

The energy difference  $\omega_0(V_\t{TP})$ depends on the TP lattice depth. For a TP lattice depth of zero, it amounts to $\omega_0(V_\t{TP}=0)=2\omega_\t{rec}=2\nicefrac{E_\t{rec}}{\hbar}$. In the limit of deep lattices, the energy difference reduces to $\omega_\t{rec}$. This is due to the lowest Bloch band becoming completely flat and therefore only the lattice along the $x$-direction contributing to the energy difference. 

The coupling strength $\lambda$ is given by $\lambda=2 M_0(V_\t{TP}) \sqrt{|V_\t{TP}U_0 N|}$, with the overlap integral $M_0$ and the dispersive shift $U_0=\nicefrac{g_0^2}{\Delta_a}$. The detuning from the atomic transition is given by $\Delta_a$ and the atom-cavity coupling strength is denoted by $g_0$. The dependence on the TP lattice depth of the overlap integral $M_0=\bra{0}\cos{(kx)}\cos{(kz)}\ket{k}$ is the result of the modification of the Bloch functions $\ket{q_z=0}$ and $\ket{q_z=k}$. Here, $q_z$ denotes the quasi-momentum along $z$ due to the presence of the TP lattice. The overlap $M_0(V_\t{TP})$ increases for increasing TP lattice depth.\\
Applying a mean-field treatment and performing an adiabatic elimination of the cavity field, and assuming that the atomic state, characterized by $\hat{\vec{J}}$, follows adiabatically the external parameters, we end up with the following expression for the TP lattice depth $V_\t{TP}$ above threshold \cite{dimer_proposed_2007}:
\begin{align}
V_\t{TP}&=\frac{\alpha}{2N^2} + \sqrt{\frac{\alpha^2}{4N^4}-\frac{\beta}{N^2}},\\
\t{with } \alpha&=\frac{(\Delta_c^2+\kappa^2)n_\t{ph}}{M_0^2(V_\t{TP})\cdot U_0}\\
\t{and } \beta&=\frac{(\Delta_c^2+\kappa^2)^2\omega_0^2(V_\t{TP})}{16 \Delta_c^2 M_0^4(V_\t{TP}) U_0^2},
\label{eq:vtp}
\end{align}
where $n_\t{ph}$ is the mean intra-cavity photon number. From this expression one can see that a decrease in atom number needs to be compensated by an increase in the TP lattice depth to maintain the same mean intra-cavity photon number. This tendency holds as well when the change in $\omega_0$ and $M_0$ due to the changing $V_\t{TP}$ is taken into account.

\subsection*{4. Atom number decay}
\begin{figure}[h!]
	\centering
	\includegraphics[width=\textwidth]{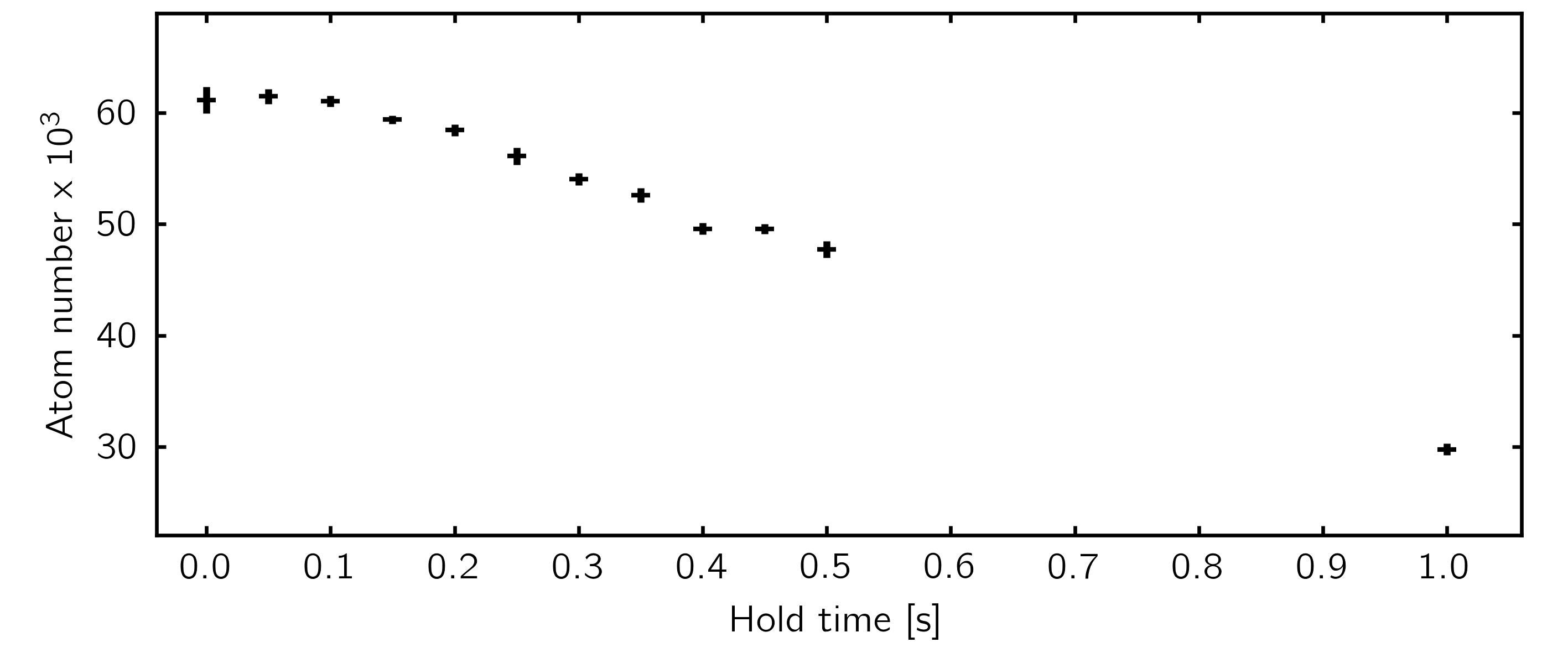}
	\caption{Atom numbers after exposing the system to the feedback stabilizing at $n_\t{ph}\approx\num{12.2}$ mean intra-cavity photons for different hold times and an adiabatic ramp down of the TP power. Data points are averages over five repetitions and error bars correspond to the standard error of the mean.}
	\label{fig:fig_si1}
\end{figure}
We measured the number of atoms in the atomic cloud after exposing the system to the feedback for different hold times. For these measurements, the feedback was stabilizing at approximately \num{12.2} mean intra-cavity photons. We initially followed the ramp protocol as described in the main text for the data shown in figure 2(c). Instead of suddenly switching off the TP lattice, we adiabatically ramped down the TP intensity with an s-shaped ramp within \SI{20}{\milli\second} in order to retrieve a harmonically trapped cloud. We then switched off all trapping potentials and let the cloud ballistically expand for \SI{8}{\milli\second}. We subsequently recorded absorption images. We fitted individual absorption images with a bimodal distribution in 2D. The resulting atom number for different hold times is shown in figure \ref{fig:fig_si1}.

We observed that the atom number stays approximately constant in the first \SI{100}{\milli\second}. For longer stabilization times up to \SI{1}{\second}, the atom number is decaying. Within the available range of data points, the atom number decay follows approximately a linear trend.

\subsection*{5. Estimating distance from critical point at start of stabilization}
We calculate the ratio $r=\nicefrac{V'_\t{TP}}{V_\t{TP,crit}}$ for each stabilized mean intra-cavity photon number shown in figure 4. Here, $V'_\t{TP}$ is the first value of the TP lattice depth where the desired photon number is reached. The critical lattice depth $V_\t{TP,crit}$ of the self-organization phase transition is extracted from fitting the photon trace with the function $n_\t{ph}(t) = o + a\cdot \t{max}(0, (V_\t{TP}(t)-V_\t{TP,crit}))^{\gamma}$, where $o$, $a$ and $\gamma$ are additional fit parameters.\\
The respective values of $n_\t{ph}$ and $r$ for all traces shown in figure 3 are given in table \ref{tab:tab1}.

\begin{table}[h!]
\center
\begin{tabular}{ |c|c|c|c|c| } 
 \hline
 $n_\t{ph}$ & \num{0.17\pm 0.04} & \num{0.3\pm 0.06}  & \num{0.58\pm 0.07} & \num{3.28\pm 0.18} \\ \hline 
 $r$ & \num{1.016\pm 0.01} & \num{1.029\pm 0.02} & \num{1.040\pm 0.025} & \num{1.051\pm 0.01} \\ \hline \hline
  $n_\t{ph}$ & \num{6.8\pm 0.2} & \num{13.7\pm 0.3} & \num{20.6\pm 0.4} & \num{27.6\pm 0.5} \\ \hline 
 $r$ & \num{1.122\pm 0.02} &  \num{1.202\pm 0.02} & \num{1.352\pm 0.01} & \num{1.507\pm 0.02}\\ \hline
\end{tabular}
\caption{Stabilized photon number $n_\t{ph}$ and ratio $r=\nicefrac{V'_\t{TP}}{V_\t{TP,crit}}$ for data shown in figure 3.}
\label{tab:tab1}
\end{table}

\bibliographystyle{iopart-num} 
\bibliography{supplement}